\documentclass[conference]{IEEEtran}
\usepackage{cite}

\usepackage{booktabs}
\usepackage{siunitx}
\usepackage{amsmath,amssymb,amsfonts}
\usepackage{algorithmic}
\usepackage{enumitem} 
\usepackage{graphicx}
\usepackage{textcomp}
\usepackage{xcolor}
\def\BibTeX{{\rm B\kern-.05em{\sc i\kern-.025em b}\kern-.08em
    T\kern-.1667em\lower.7ex\hbox{E}\kern-.125emX}}
\begin{document}

\title{Compressive Beam Alignment for Indoor Millimeter-Wave Systems}

\makeatletter
\newcommand{\linebreakand}{%
  \end{@IEEEauthorhalign}
  \hfill\mbox{}\par
  \mbox{}\hfill\begin{@IEEEauthorhalign}
}

 \author{\IEEEauthorblockN{April Junio,  Rafaela Lomboy$^*$, Raj Sai Sohel Bandari, and Mohammed E. Eltayeb}
 \IEEEauthorblockA{\textit{Department of Electrical \& Electronic Engineering} \\
 \textit{$^*$Department of Computer Science} \\
 \textit{California State University, Sacramento, Sacramento, USA} \\
 Email: \{apriljunio, rlomboy, rbandari, mohammed.eltayeb\}@csus.edu}
 \vspace{-0.25in}
 }

%
\maketitle

\begin{abstract}
The dynamic nature of indoor environments poses unique challenges for next-generation millimeter-wave (mm-wave) connectivity. These challenges arise from blockages due to mobile obstacles,  mm-wave signal scattering caused by indoor surfaces,  and imperfections in the users phased antenna arrays.  Traditional compressed sensing (CS) based beam alignment techniques enable swift mm-wave connectivity with a limited number of measurements.  These techniques, however,  rely on prior knowledge of the communication channel model and the user's array manifold to design the sensing matrix and minimize angle quantization errors.  This limits their effectiveness in dynamic environments.  This paper proposes a novel CS-based beam alignment technique for mm-wave systems operating in indoor environments.  Unlike prior work that rely on  knowledge of the user's antenna architecture, communication codebook, and channel, the proposed technique is agnostic to these factors.  The proposed formulation eliminates angle quantization errors by mapping the recovered angular directions directly onto the user's specific codebook. This is achieved by exploiting the energy compaction property of the Discrete Cosine Transform (DCT) to compress and identify the strongest cluster locations in the transform domain for robust beamforming.  Experimental results at 60 GHz demonstrate successful recovery of the mm-wave power distribution in the angular domain, facilitating accurate beam alignment with limited measurements when compared to exhaustive search solutions.

\end{abstract}

\begin{IEEEkeywords}
Millimeter-wave communications,  indoor beam alignment, compressed sensing.
\end{IEEEkeywords}

\section{Introduction}

The demand for low latency and high-speed communications to support new technology and applications such as advanced telehealth,  immersive reality experiences, and high-definition video streaming is on the rise \cite{p2}.  The abundance of bandwidth and high data rates in the millimeter-wave (mm-wave) band enables service providers to meet those demands \cite{p3}.  However promising, the deployment of mm-wave communication systems is challenged by the sensitivity to line-of-sight (LoS) link blockages and requires precise alignment of the communication beams.  While existing techniques such as codebook-based hierarchical beam training and compressed channel estimation have been proposed to address these challenges  \cite{p4, p5,p51,b1,p6},  they exhibit limitations in dense indoor environments where intermittent blockages are imminent.  Furthermore,   these techniques require knowledge of the complete array response of all receivers,   do not account for potential distortions introduced by the antenna front-end hardware,  and  often rely on a pre-defined channel model, which might not accurately reflect real-world conditions.  To address these limitations,  novel methods are required that can mitigate mm-wave link blockages by swiftly identifying alternate communication paths and ensure robust performance in dynamic indoor environments regardless of the receiver's antenna configuration.

Compressed sensing (CS) has emerged as a promising signal processing technique for swift mm-wave channel estimation due to its ability to exploit channel sparsity \cite{p11,p12,p13}.   The core principle of CS dictates that a sparse signal can be accurately reconstructed from a limited set of measurements using optimization algorithms, either in its original domain or a transformed domain \cite{p11}. Application of CS in mm-wave channel estimation translates to efficient bandwidth utilization and faster channel acquisition compared to traditional beam alignment methods  \cite{p12}.  Nonetheless,  existing CS-based mm-wave beam alignment techniques primarily focus on outdoor channels \cite{b1,p6,p12,  p13}.  Outdoor environments are generally static,  and mm-wave propagation channels exhibit spatial sparsity  \cite{p13}.  This sparsity aligns well with the assumptions of compressed sensing. Conversely,  indoor environments present a unique challenge due to the presence of numerous objects that significantly affect signal propagation, resulting in increased scattering and potentially less sparse channels. Fig.~\ref{fig:fig1} illustrates this by plotting the received signal strength (RSS) measurements obtained at a line-of-sight (LoS) receiver within an indoor environment. As expected, the figure reveals that the strongest power is received at the LoS angles (0° for both transmit and receive). However, it also demonstrates the presence of significant scattered energy at non-line-of-sight (NLoS) directions. These secondary peaks likely arise from multipath propagation due to reflections and scattering induced by the interaction of the mm-wave signal with surrounding objects in addition to energy leakage from the antenna sidelobes.   This scattering profile underscores the key difference between indoor and outdoor mm-wave channels.

This paper presents a novel CS approach for beam alignment in indoor mm-wave environments.  We depart from prior CS-based methods which rely on prior knowledge of the communication channel model and array manifold.   Instead, we introduce a framework that leverages the discrete cosine transform (DCT) domain to achieve sparsity and approximate the directions of the strongest channel clusters (power) in the transform domain.  This framework offers two key advantages: (i) model-agnostic operation as it eliminates the need for prior knowledge of the channel model, and ii)  device-specific operation by directly mapping angles-of-arrival (AoA) and/or angles-of-departure (AoD) to the user's quantized codebook.   This is achieved by sampling power measurements at random directions and exploiting the energy compaction property of the DCT to recover a compressed representation (see Fig.  \ref{fig:dct}) that closely approximates the original spatial domain power distribution.  This behavior is analogous to a low-pass filter in image processing, which blurs the image but can reconstruct missing information to some extent.  To the best knowledge of the authors, this work represents the first implementation of CS-based beam alignment in a realistic indoor dense environment.

The rest of the paper is organized as follows.  Sec. II introduces the system model and problem formulation, 
  Sec.  III presents our proposed compressed mm-wave beam alignment solution,  setup and methodology.  Sec. IV presents the measurement setup and methodology.  Our experimental results are presented and discussed  in V.  Finally,  Sec. VI concludes the paper.

\section{System Model and Problem Formulation}
\subsection{System Model}
We consider a system where a stationary mm-wave access point (AP) with $t$ antennas communicates with a stationary single receiver equipped with $r$ antennas.  The access point uses one of $p$  beamforming vectors present in its codebook   $\boldsymbol{\mathcal{F}} = \{ \mathbf{f}_1, \mathbf{f}_2, ..., \mathbf{f}_p$\},  and similarly, the receiver uses one of its $q$ beamforming vectors $\boldsymbol{\mathcal{W}} = \{\mathbf{w}_1, \mathbf{w}_2, ...,  \mathbf{w}_q\}$ for communication.  The beamforming vector $\mathbf{f}_i \in \mathcal{C}^{t\times 1}$ steers the transmit beam towards the angle $\theta_i$, and similarly, the combining vector  $\mathbf{w}_j \in \mathcal{C}^{r\times 1}$  at the receiver steers the receiver's beam towards the angle $\theta_j$.  Let $s$,  $\mathbb{E} [ |s|^2 ]=1$,  be the complex symbol transmitted by the AP to the receiver using the beamformer $\mathbf{f}_j$,  the received signal at the receiver using the combing vector  $\mathbf{w}_i$ is given by
\begin{eqnarray}\label{eq:s1}
y = \mathbf{w}^*_i \mathbf{H} \mathbf{f}_j s + {e},
\end{eqnarray}
where $i=1,2,.., q$,  $j=1,2,...,p$,  and $\mathbf{e} \sim \mathcal{CN}(0, \sigma^2)$ represents the additive white Gaussian noise  with a complex normal distribution. The matrix  $\mathbf{H}$ of size $r \times t$ 
 represents the unknown mm-wave channel between the AP and the receiver.

\subsection{Problem Formulation}
In this paper,  the received power at the receiver is adopted as the performance metric. Therefore,  optimal transmit and receive beam selection occurs when the AP and the receiver pick the beamforming and combining vectors  that maximizes the received power at the receiver,  i.e.  $\mathbb{E} [ |y|^2 ]$.  This is achieved by selecting the ideal transmit beamforming vector $\mathbf{f}^{\star}$ and receive combining vector $\mathbf{w}^{\star}$ as follows
\begin{eqnarray}\label{eq:s2}
<\mathbf{f}^{\star},  \mathbf{w}^{\star}> =  \arg \max_{{{ \mathbf{w} \in \boldsymbol{\mathcal{W}},  \mathbf{f}\in \boldsymbol{\mathcal{F}}}}} |\mathbf{w}^* \mathbf{H} \mathbf{f}|^2.
\end{eqnarray}
Without explicit knowledge of the channel $\mathbf{H}$, the optimal transmit and receive vectors can be obtained via an exhaustive search encompassing all codebook entries. However, this approach suffers from high computational complexity, making it impractical for real-world scenarios.  In the next section, we propose a method  that  identifies the angular distribution of the strongest channel clusters  without the need for an exhaustive search process.

\begin{figure}
\center
 \includegraphics[width=7cm]{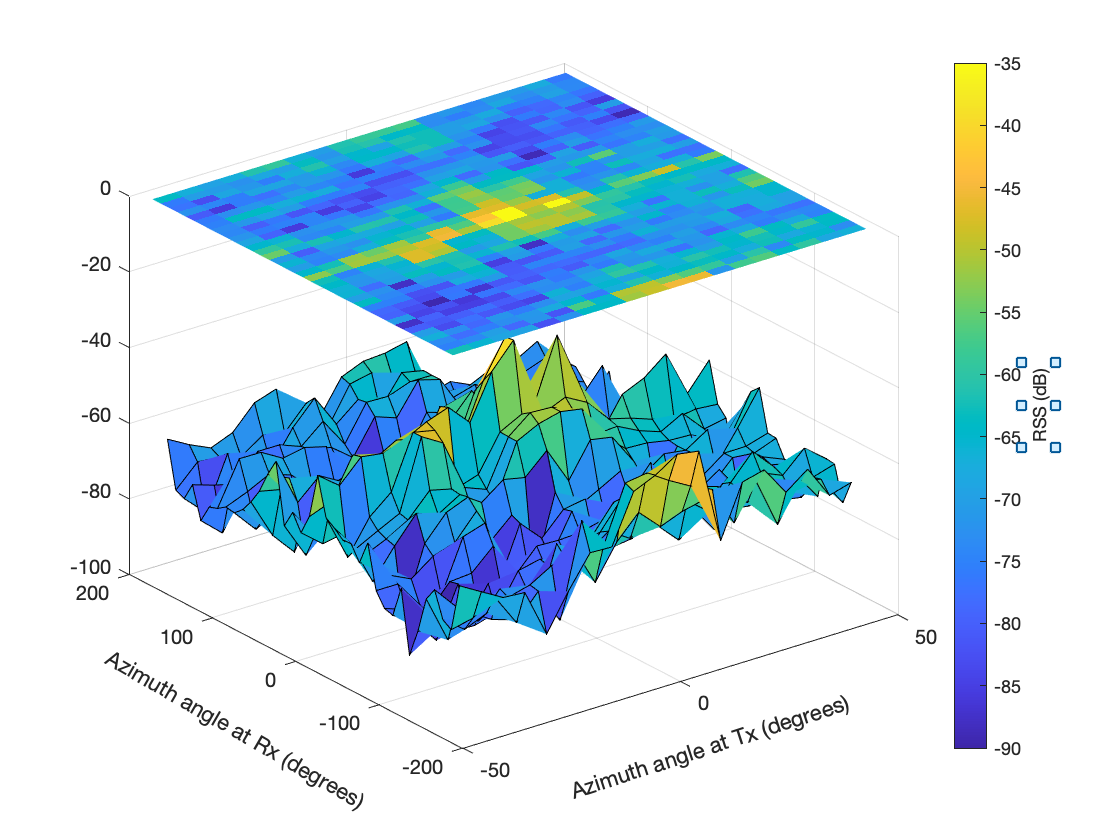}
\caption{ Received signal strength (RSS) measurements as a function of beam orientation for a 16-element phased antenna array operating at 60 GHz in a 58 square meter indoor laboratory environment.  
}
\label{fig:fig1}
\end{figure}

\begin{figure}
\center
 \includegraphics[width=7cm]{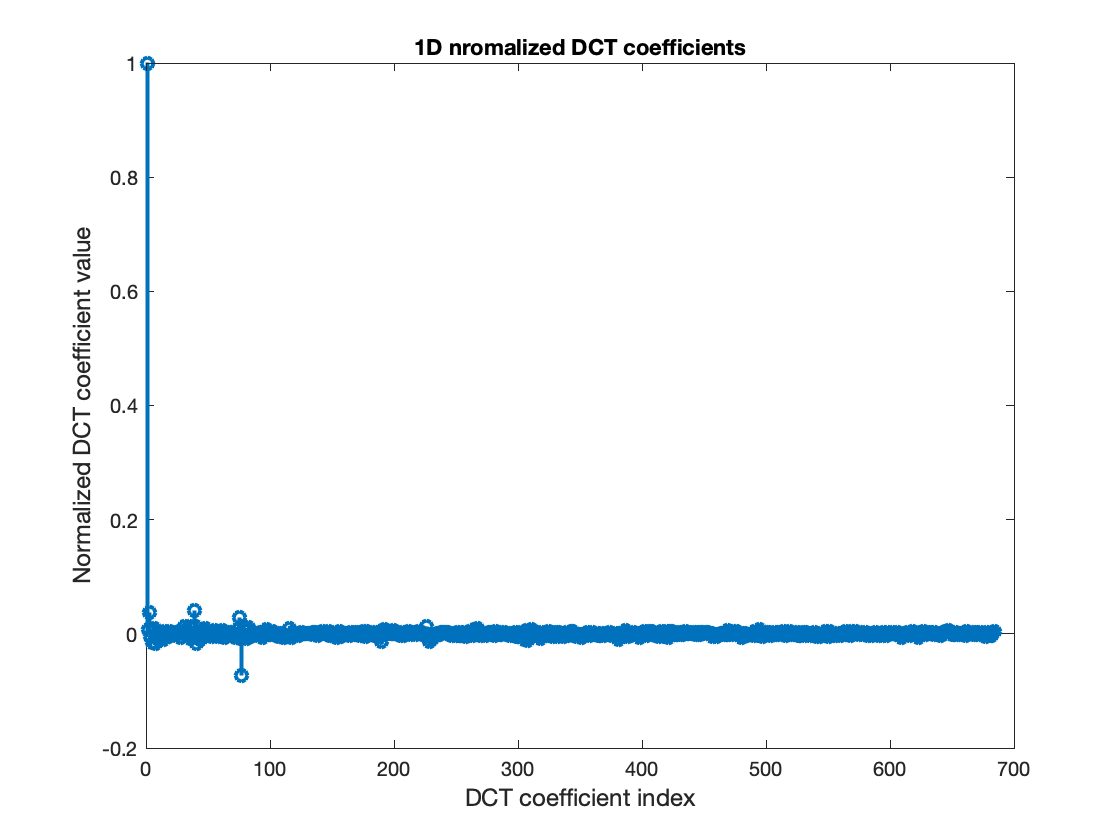}
\caption{ Normalized DCT coefficients of the RSS measurements shown in Fig.  \ref{fig:fig1}.  The DCT efficiently concentrates signal energy in low-frequency coefficients,  resulting in a sparse representation which we exploit in this paper.}
\label{fig:dct}
\end{figure}

\section{Proposed Compressed MM-Wave Beam Alignment Solution}
We present a new method to determine the received power,  referred to as the received signal strength (RSS) in this paper, across all possible combinations of the transmit and receive beamforming vectors.  It then identifies the optimal beam vectors ($\mathbf{f}^{\star},  \mathbf{w}^{\star}$) that maximize the RSS at the receiver without explicitly estimating the channel matrix $\mathbf{H}$.  The following subsections will delve into the details of this approach.

\subsection {Initial Beam Measurements}
To initiate sensing, the transmitter transmits $m_1$ beams using randomly selected $m_1$ beamforming vectors from its codebook $\boldsymbol{\mathcal{F}}$.  Similarly,   the receiver uses randomly selects $m_2$ combining vectors from its codebook $\boldsymbol{\mathcal{W}}$ for each transmit beam to capture projections of the spatial distribution of the signal strength as follows
\begin{eqnarray}\label{eq:s1}
\mathbf{Y} = |\mathbf{W}_{m_2}^* \mathbf{H} \mathbf{F}_{m_1}  +  \mathbf{E}|^2.
\end{eqnarray}
The matrix $\mathbf{Y} \in \mathbb{C}^{m_2 \times m_1}$ contains the sampled signal strength measurements using $m=m_1 m_2$ transmit and receive beams,  the combing matrix $\mathbf{W}_{m_2} \in \mathbb{C}^{r \times m_2}$  consists of $m_2$ combining vectors randomly selected from the codebook $\boldsymbol{\mathcal{W}}$,   the beamforming matrix $\mathbf{F}_{m_1} \in \mathbb{C}^{t \times m_1}$  consists of $m_1$ beamforming vectors randomly selected from the codebook $\boldsymbol{\mathcal{F}}$,  and $ \mathbf{E}\in \mathbb{C}^{{m_2 \times m_1}}$ is the complex additive noise matrix.

\subsection{Sparse formulation for recovering missing measurements}
 For ease of exposition,  we omit the effect of noise and rewrite (\ref{eq:s1}) as
\begin{eqnarray}\label{eq:s2}
\tilde{\mathbf{Y}} &=& |\mathbf{W}_{m_2}^* \mathbf{H} \mathbf{F}_{m_1}|^2 = |\mathbf{S}_2 \mathbf{W}_q^* \mathbf{H} \mathbf{F}_p\mathbf{S}_1|^2\\ \label{eq:s25}
&=&  \mathbf{S}_2 |  \mathbf{W}_p^* \mathbf{H} \mathbf{F}_q|^2\mathbf{S}_1 = \mathbf{S}_2\mathbf{\Phi} \mathbf{S}_1.
\end{eqnarray}
In  (\ref{eq:s2}),  we expressed $\mathbf{W}^*_{m_2}$ as $\mathbf{S}_2 \mathbf{W}_q^*$ and $\mathbf{F}_{m_1}$ as  $\mathbf{F}_p\mathbf{S}_1$.  The combining matrix $\mathbf{W}_q$ consists of of all the combining vectors in  $ \boldsymbol{\mathcal{W}}$,  and the beamforming  matrix $\mathbf{F}_p$ consists of all the vectors in  $ \boldsymbol{\mathcal{F}}$.  The random selection matrices  $\mathbf{S}_1 \in  \mathbb{R}^{p \times m_1} $ and   $\mathbf{S}_2 \in \mathbb{R}^{m_2 \times q}$ are binary matrices where each row has a cardinality (number of ``1"s) of 1, and each column has a cardinality of at most 1.  The unknown matrix $ \mathbf{\Phi} \ = |  \mathbf{W}_p^* \mathbf{H} \mathbf{F}_q|^2$ is of size $p \times q$ and carries the projections of mm-wave signal strength across all transmit and receive codebooks entries. This matrix is  unknown and can be estimated via exhaustive search over all transmit and receive beams.  This brute force approach necessitates a total of $n=pq$ measurements,  where $p$ and $q$ represent the dimensions of the transmit and receive codebooks, respectively. 

Vectorizing the matrix $\tilde{\mathbf{Y}}$ in (\ref{eq:s25}) yields
\begin{eqnarray}\label{eq:s3}
  \tilde{\mathbf{y}} =   \underbrace{\mathbf{S}_2^T  \otimes  \mathbf{S}_1}_{\mathbf{A}} \underbrace{\text{Vec} (\mathbf{\Phi})}_{\mathbf{x}} ,
\end{eqnarray} 
where   $\tilde{\mathbf{y}} = \text{Vec}(\tilde{\mathbf{Y}} ) $  and $\otimes$ is the Kronecker product operator.  While the matrix $\mathbf{\Phi}$ might not be inherently sparse in the spatial domain,  we exploit the energy compaction property of the DCT to enforce $\mathbf{\Phi}$ be sparse in the DCT domain.  The DCT has the property of concentrating the signal's energy into a few coefficients, particularly the low-frequency ones. By neglecting the higher-frequency DCT coefficients of $\mathbf{\Phi}$, we enforce sparsity in the transformed domain.  Using compressed sensing,  an approximation to the DCT of $\mathbf{\Phi}$ is obtained and used to compute an approximation of  $\mathbf{\Phi}$ in the spatial domain.

\begin{figure}
\center
 \includegraphics[width=6cm]{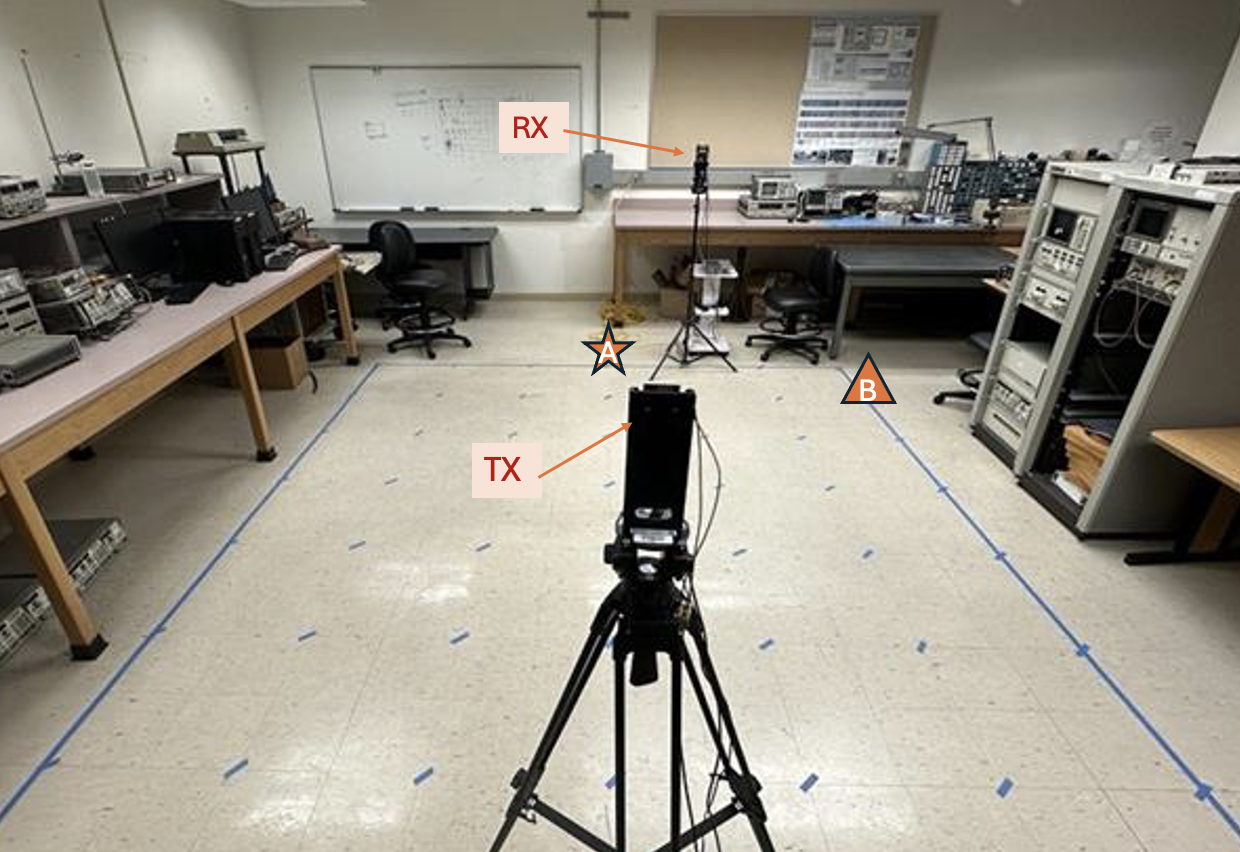}
\caption{ View of the mm-wave propagation environment.  The transmit and receive antennas were situated 4.3 meters (Location A) and 4.5 meters (Location B) apart.  Antennas heights are set to 1.6 meters. }
\label{fig:lab}
\end{figure}

\begin{figure}
 \includegraphics[width=8cm]{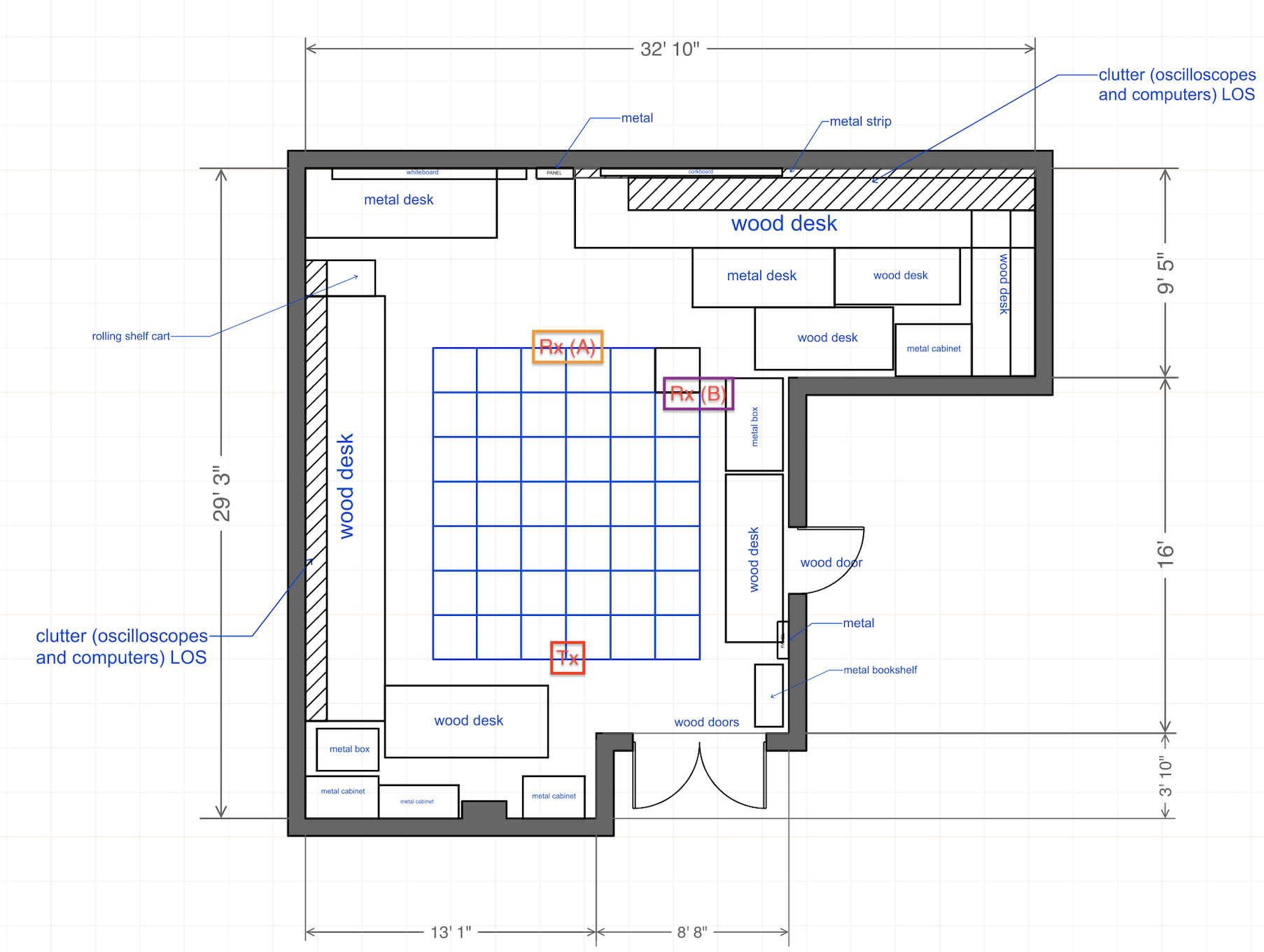}
\caption{ Map and dimensions of the  indoor environment.}
\label{fig:lab2}
\end{figure}

\subsection{Compressed sensing recovery}
To recover $\mathbf{x}$ in the transform domain rewrite (\ref{eq:s3})  as follows
\begin{eqnarray}\label{eq:s4}
  \tilde{\mathbf{y}} =   \mathbf{Ax} = \mathbf {A} \mathbf{\Psi}^{-1} \hat{\mathbf{x}}.
\end{eqnarray}
In (\ref{eq:s4}),  $\mathbf{\Psi}^{-1}$ represent the inverse DCT  matrix,  and $\hat{\mathbf{x}}$ carries the DCT coefficients of the power measurement vector ${\mathbf{x}} = \text{Vec} (\mathbf{\Phi})$.    To recover the vector $\hat{\mathbf{x}}$,  we perform CS recovery by solving the following $\ell_0$-minimization problem as follows
 \begin{eqnarray}\label{eq:s5}
 \min \| \hat{\mathbf{x}} \|_0 \quad  \text{s.t. }   \tilde{ \mathbf {y}} \mathbf{ = A} \mathbf{\Psi}^{-1} \mathbf{ \hat{\mathbf{x}}}.
\end{eqnarray}
While there are many different methods used to solve sparse approximation problems (see, e.g.,  \cite{ Ncs, lasso} ), we employ the least absolute shrinkage and selection operator (LASSO) \cite{lasso} to recover the entries of $\mathbf{ \hat{\mathbf{x}}}$.   We adopt the LASSO since it does not require prior knowledge of the support of the vector $\mathbf{ \hat{\mathbf{x}}}$.  This makes it a suitable detection technique in the absence of a channel model.  The  LASSO estimate of (\ref{eq:s5}) is given by \cite{lasso}
\begin{eqnarray}\label{eq:s4l}
\arg \min_{\mathbf{ \hat{\mathbf{x}}}  \in \mathbb{R}^{n\times 1}}\frac{1}{2} \| \mathbf{y} - \mathbf{A} \mathbf{\Psi}^{-1} \mathbf{ \hat{\mathbf{x}}} \|_2^2 + \Omega \sigma \|  \mathbf{ \hat{\mathbf{x}}}\|_1,
\end{eqnarray}
where $\sigma$ is the standard derivation of the noise,  and $\Omega$ is a regularization parameter.  After recovery,  the entires of $\mathbf{ \hat{\mathbf{x}}}$ are rearranged to obtain an estimate of the power distribution matrix $\hat{\mathbf{\Phi}}$.

\subsection{Optimal beam selection}
The indices of optimal beamforming/combining vector pair ($i^{\star},  j^{\star}$)  that correspond to the maximum estimated received signal strength is selected as follows
 \begin{eqnarray} \nonumber
 <i^\star, j^\star >  =   \arg \max_{{{ i,j }}} \hat{\boldsymbol{\Phi}}_{i,j}, 
\end{eqnarray}
where $i = 1,...,p,  \text{ and  } j= 1,...,q.$ The selected transmit beamforming vector is $\mathbf{f}^{\star} = \boldsymbol{\mathcal{F}}_{:,i^\star}$ and the selected receive combining vector is $\mathbf{w}^{\star} = \boldsymbol{\mathcal{W}}_{:,j^\star}$

\begin{figure}
\center
 \includegraphics[width=8.7cm]{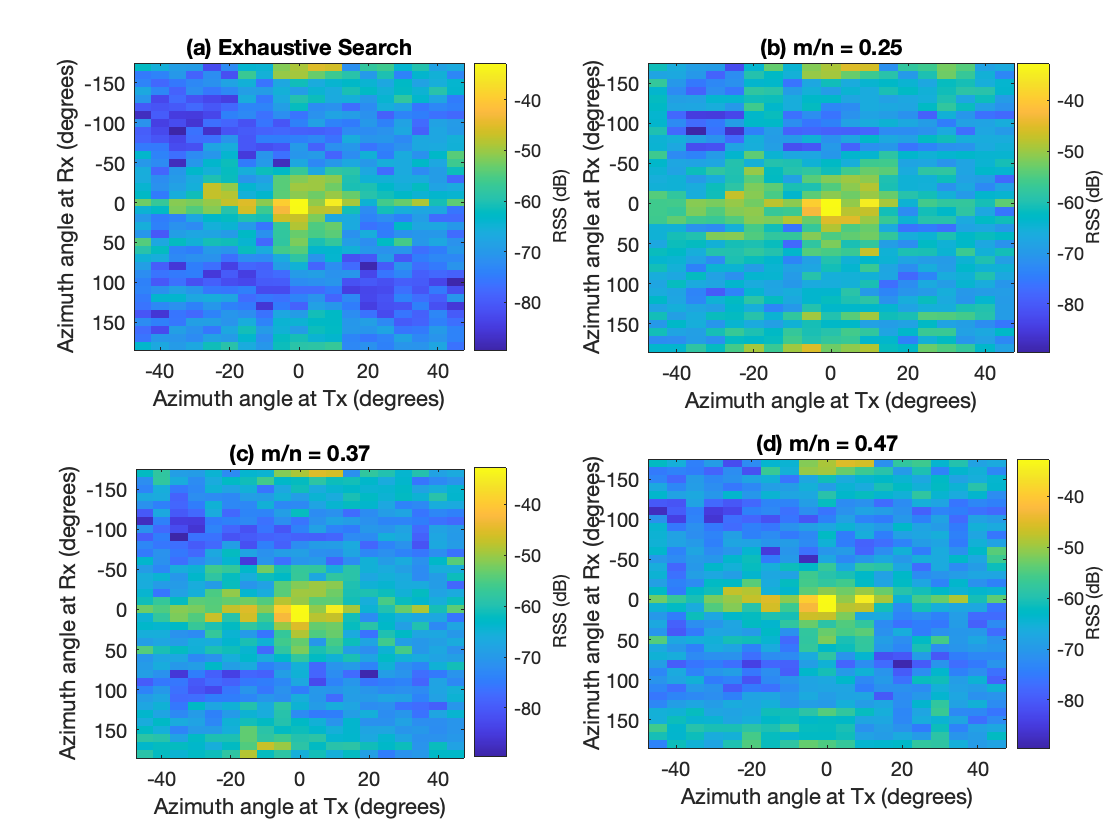}
\caption{RSS versus the TX/RX beam orientation at location A when using (i) exhaustive search over all TX/RX angles (top left figure),  (ii) CS recovery with ($m/n = 0.25$),  (iii) CS recovery with ($m/n = 0.37$),  and (iv) CS recovery with  ($m/n = 0.47$).}
\label{fig:fig4}
\end{figure}

\section{ Measurement Setup and Methodology}\label{ms}
Indoor measurements were conducted in a lab environment with surfaces of varying materials including wood panel,  glossy white-board, metal cabinets, and other environmental clutter as shown in Fig.  \ref{fig:lab}.  The transmitter (TX) was positioned at a fixed location,  while the receiver (RX) was positioned at two locations,  namely location A and B as shown in Fig. \ref{fig:lab2}.   At Location A, the receiver is aligned with the transmitter to create a LoS link.  At Location B,  the receiver is situated adjacent to a metallic equipment rack thereby resulting in an additional anticipated cluster as shown in Fig. \ref{fig:fig5}.

\subsection{Phased antenna array setup}
Two 60 GHz mm-wave  phased antenna array transceiver kits (Sivers EVK02001) \cite{SiversIMA} were used for measurements as depicted in Fig. \ref{fig:lab}.   The beam direction can be electronically steered from +45 to -45 degrees in the azimuth plane.   Each antenna kit includes a 16 element antenna patch antenna module that is steered using codebooks  $\boldsymbol{\mathcal{F,W}}$.   The transmit and receive  beams can be steered in the azimuth plane in the range $-45^o$ to $45^o$.  The transmit and receive codebooks can be found in \cite{Deepsense}.   

\subsection{Measurement procedure}
A universal software radio peripheral  (USRP) is used to generate a sinusoidal baseband signal with a sampling rate of 1 MHz.  The phased antenna array transceiver kit up converts this signal to 60 GHz  before transmission to the receiver.  The receiver kit and USRP downconvert the received signal back to baseband.  The power spectrum of the down converted signal is computed using a flattop window filter.   To capture signal strength measurements across various angles,   the receiver is steered in the azimuth plane in the range  $-180^o$,  to $180^o$ in increments of $10^o$.  Similarly, the transmit angles are electronically steered from +45 to -45 degrees in increments of 5 degrees in the azimuth plan.  The received signal strength is extracted from the FFT power spectrum for each receive angle.

\begin{figure}
\center
 \includegraphics[width=9cm]{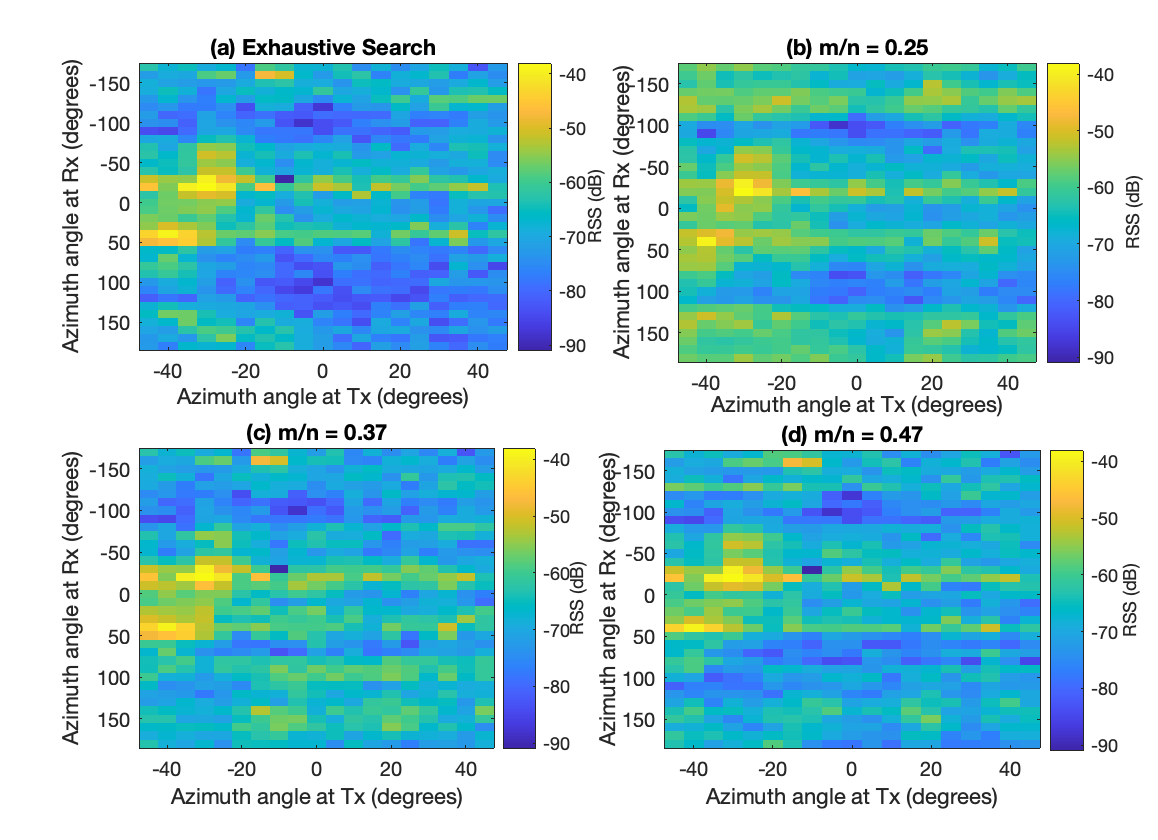}
\caption{RSS versus the TX/RX beam orientation at location B when using (i) exhaustive search over all TX/RX angles (top left figure),  (ii) CS recovery with ($m/n = 0.25$),  (iii) CS recovery with ($m/n = 0.37$),  and (iv) CS recovery with  ($m/n = 0.47$).}
\label{fig:fig5} 
\end{figure}

\section{Experimental Results and Discussions}
In this section, we use real-world data to test the performance of the proposed compressed beam alignment technique outlined in Sec.  III using the measurement setup outlined in Sec. \ref{ms}.  We analyze the efficacy of the proposed  beam alignment technique based on measured RSS.  The transmit codebook $\boldsymbol{\mathcal{F}}$  consists of 19 beam steering vectors,  while the receive codebook $\boldsymbol{\mathcal{W}}$ consists of 36 entries. 

In Fig. \ref{fig:fig4} random samples were taken at Location A and the proposed CS technique is applied to recover the  power distribution across the complete angular domain.  When the number of measurements/samples are 171, corresponding to a ratio of $\frac{171}{684} = 0.25$,   the proposed technique was able to identify  the location of the strong cluster, however,  we see also observe some false positives, ie power at random angles, when compared to the exhaustive search method.  We also observe that these false positives diminish for higher number of measurements.  This trend is observed at Location B as well as shown in Fig.  \ref{fig:fig5}.

To account for variability and evaluate performance under different beam selection conditions, 1000 iterations of random  codebook selection were conducted to generate an average performance metric in Figs.  \ref{fig:fig6}, and \ref{fig:fig7}.   Fig.   \ref{fig:fig6} shows that the normalized mean square error (NMSE),   defined as  $\frac{\|{\mathbf{\Phi}}-\hat{\mathbf{\Phi}}\|_2^2}{\|{\mathbf{\Phi}}\|_2^2}$, to calculate the  average power distribution reconstruction error.  The figure shows that the NMSE decreases  with an increasing number of compressed sensing measurements at both locations (A and B). This results since more measurements lead to fewer false positives.   In Fig.  \ref{fig:fig7}, we plot the average RSS loss for the best beam direction, which is the difference between the highest RSS obtained via exhaustive search and that achieved using the proposed technique,  versus the fraction of measurements.  The figure shows that with a fraction of 0.47 measurements,  the average loss between the selected RSS achieved via exhaustive search and the proposed technique is  less than 1 dB  at location A and close to 1.6 dB at Location B.   Location B experience higher loss due to the presence scatters which results in additional coefficients in the transform domain.  Nonetheless, the fiure shows that the RSS loss decreases as the number of measurements increases for both locations.

%

\section{Conclusions}
In this paper, we proposed and evaluated a novel mm-wave beam alignment technique specifically suited for dense indoor environments.   Unlike conventional methods that exhaustively scan the signal strength across all directions,  the proposed technique scans a fraction of the transmit-receive directions and leverages the energy compaction property of the DCT to recover the direction of the strongest clusters in the transform domain.  Consequently,  the optimal transmit-receive angles can be estimated.  Our results demonstrate that the proposed approach significantly reduces the beam alignment  time compared to exhaustive search methods,  thus making it highly attractive for practical mm-wave deployments in dense indoor environments.  Future work will focus on beam alignment for multi-user scenarios and leveraging environment information to further reduce the fraction of samples required for  beam alignment.

\begin{figure}
\center
 \includegraphics[width=8.25cm]{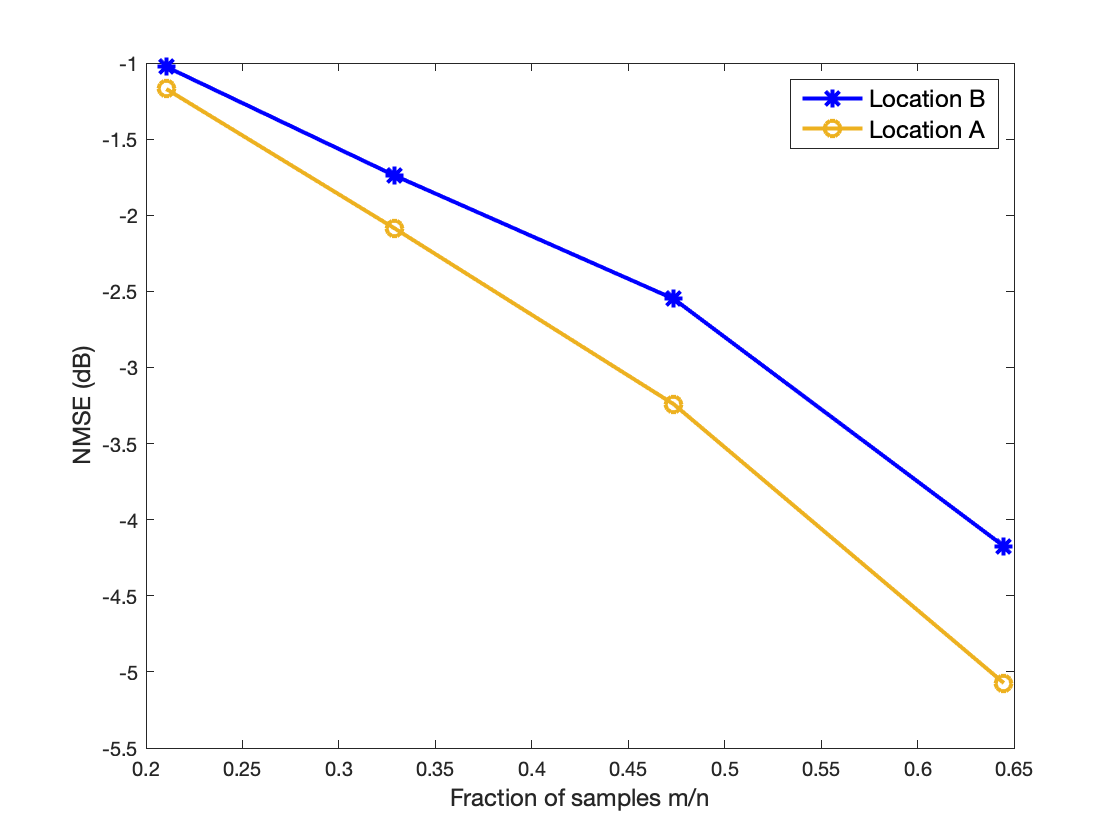}
\caption{  NMSE of the recovered power distribution versus the fraction of CS measurements (number of measurements normalized by the total number of search directions) at locations A and B.}
\label{fig:fig6}
\end{figure}

\begin{figure}
\center
 \includegraphics[width=8.25cm]{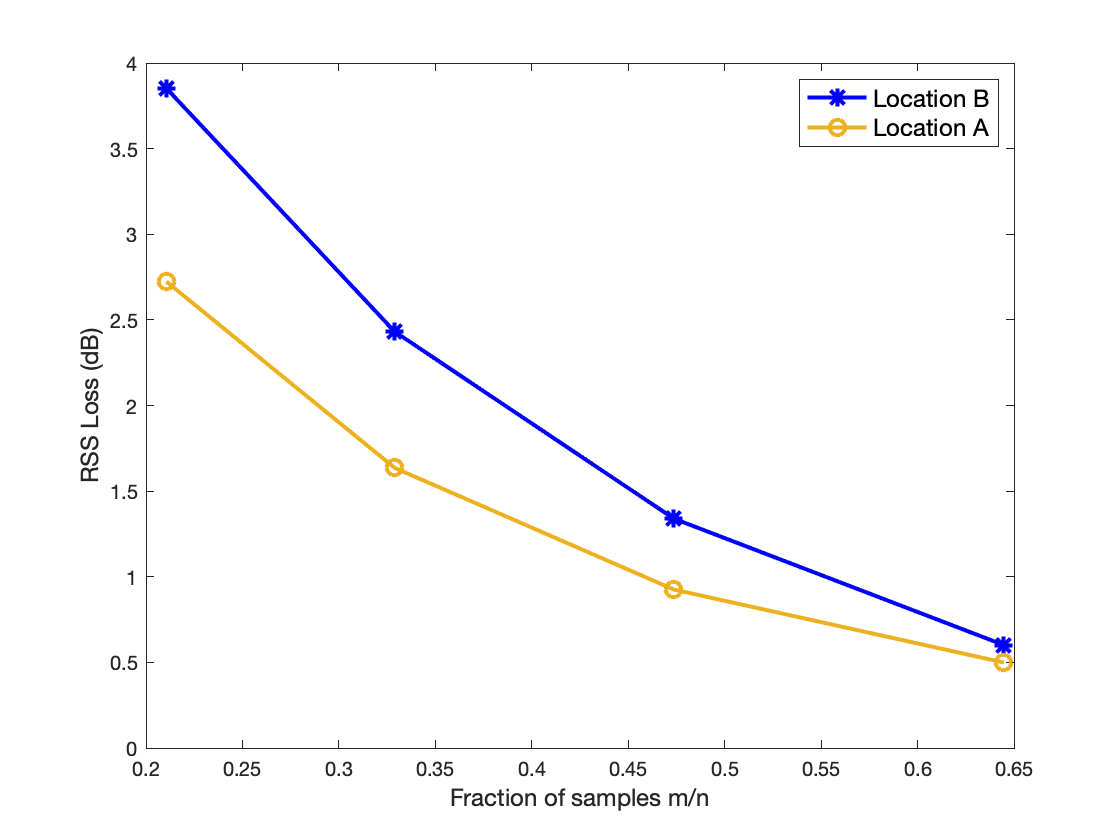}
\caption{Best beam RSS loss versus the fraction of CS measurements (number of measurements normalized by the total number of search directions) at Locations A and B.  The RSS loss decreases with increasing number of CS measurements.} 
\label{fig:fig7}
\end{figure}

\section*{Acknowledgment}
This material is based on work supported by the National Science Foundation under grant No. NSF-2243089.


\end{document}